\documentclass[twoside]{article}
\usepackage{fleqn,espcrc2}

\usepackage{graphicx}

\newcommand{\AmS}{{\protect\the\textfont2
  A\kern-.1667em\lower.5ex\hbox{M}\kern-.125emS}}

\title{Magnetotransport study of the charged stripes in high-$T_c$ cuprates}

\author{Yoichi Ando,\address{Central Research Institute of 
Electric Power Industry (CRIEPI), Komae, Tokyo 201-8511, Japan}
A. N. Lavrov,$^{\rm{a}}$\thanks{On leave from Institute of Inorganic Chemistry, 
630090 Novosibirsk, Russia.  A. N. L. acknowledges support from JISTEC.}
and Kouji Segawa$^{\rm{a}}$}

\begin{document}

\begin{abstract}
We present a study of the in-plane and out-of-plane magnetoresistance (MR) 
in heavily-underdoped, antiferromagnetic YBa$_2$Cu$_3$O$_{6+x}$, 
which reveals a variety of striking features. 
The in-plane MR demonstrates a ``$d$-wave"-like anisotropy 
upon rotating the magnetic field $H$ within the $ab$ plane. 
With decreasing temperature below 20-25 K, the system acquires memory: 
exposing a crystal to the magnetic field results in a persistent 
in-plane resistivity anisotropy. 
The overall features can be explained by assuming that 
the CuO$_2$ planes contain a developed array of stripes 
accommodating the doped holes, and that the MR is associated 
with the field-induced topological ordering of the stripes.
\vspace{1pc}
\end{abstract}

\maketitle

\section{INTRODUCTION}

The conducting state of the high-$T_c$ cuprates appears as a result of 
hole or electron doping of the parent antiferromagnetic (AF) insulator. 
In general, there is a tendency of doped holes in the AF environment 
to phase-segregate, which may give rise to an 
intriguing microscopic state with carriers gathered within 
an array of quasi-1D ``stripes" separating AF domains \cite{Emery,Nagaev,Borsa}. 
An ordered striped structure has been observed in 
La$_2$NiO$_{4.125}$ (Ref. \cite{Tranquada1}) 
and in La$_{1.6-x}$Nd$_{0.4}$Sr$_x$CuO$_4$ (Nd-LSCO, Ref. \cite{Tranquada2}), 
while most superconducting cuprates demonstrate 
incommensurate magnetic fluctuations \cite{Yamada} which can be 
considered as {\it dynamical} stripe correlations \cite{Emery,Yamada}. 
The dynamical stripes might be responsible for 
the peculiar normal state of cuprates as well as for the 
occurrence of superconductivity \cite{Emery}, but still very little 
is known about the electron dynamics in the stripes.

In this paper, we report an extraordinary behavior of the 
magnetoresistance (MR) in antiferromagnetic YBa$_2$Cu$_3$O$_{6+x}$
(YBCO), which provides evidences that conducting stripes actually 
exist in CuO$_2$ planes and have a considerable impact on 
the electron transport.  
We note that a recent study of the Hall effect in Nd-LSCO 
also provides insights into the electron transport in static stripes
\cite{Noda}.

\section{EXPERIMENTAL}

The high-quality YBCO single crystals were grown by the flux method in 
Y$_2$O$_3$ crucibles, and a high-temperature annealing was used 
to reduce their oxygen content. 
The MR was measured by sweeping the magnetic field at fixed temperatures 
stabilized by a capacitance sensor with an accuracy of $\sim$1 mK. 
The angular dependence of the MR was determined by rotating the sample 
within a 100$^{\circ}$ range under constant magnetic fields up to 16 T.

\begin{figure}[tb]
\vspace{-10mm}
\includegraphics[scale=0.42]{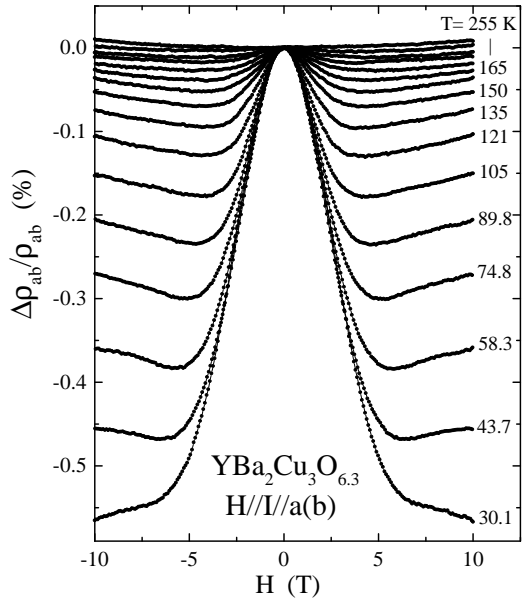}
\vspace{-40mm}
\caption{Longitudinal in-plane MR of YBCO in the antiferromagnetic
composition. 
The data are averaged over several field sweeps.}  
\label{fig1} 
\end{figure}

\section{RESULTS}

The heavily-underdoped YBCO crystals, 
though located deep in the AF range of the 
phase diagram ($x$$\approx$0.3), are far from conventional insulators: 
the in-plane resistivity $\rho_{ab}$ remains ``metallic" at high $T$ 
and it grows slower than expected for the 
hopping electron transport at low $T$ \cite{Lavrov,stripe}. 
These AF crystals demonstrate an unusual behavior of the in-plane MR, 
$\Delta\rho_{ab}/\rho_{ab}$, when the magnetic field $H$ is applied 
along the CuO$_2$ planes, as shown in Fig. 1. 
At weak fields, the longitudinal in-plane MR [$H$$\parallel$$I$$\parallel$$ab$] 
is negative and follows roughly a $T$-independent $\zeta H^2$ curve, 
and then abruptly saturates above some threshold field. 
The threshold field and the saturated MR value gradually increase 
with decreasing temperature. 
The MR anomaly becomes noticeable near the N\'eel temperature 
$T_N$$\approx$230 K ($T_N$ can be obtained from the $\rho_c(T)$ data, 
as reported in Ref. \cite{Lavrov,MR}), 
but evolves rather smoothly through $T_N$, 
which indicates that the long-range AF order itself is not 
responsible for its origin \cite{stripe}.

When the magnetic field is turned in the plane to become perpendicular 
to the current [$H$$\parallel$$ab$; $H$$\perp$$I$], 
the low-field MR term just switches its sign, 
retaining its magnitude and the threshold-field value \cite{stripe}. 
This can be graphically shown in the MR data taken upon rotating 
$H$ within the $ab$ plane, which revealed a striking anisotropy with a 
``$d$-wave"-like symmetry; 
i.e. $\Delta\rho_{ab}/\rho_{ab}$ changes from negative at $\alpha$=0$^{\circ}$ 
to positive at $\alpha$=90$^{\circ}$, being zero at about 45$^{\circ}$
($\alpha$ is the angle between $H$ and $I$), see Fig. 2. 
It is worth noting that the low-field MR feature is not observed at all 
when the magnetic field is applied along the $c$-axis. 

The most intriguing peculiarity of the low-field MR appears 
at temperatures below $\sim$25 K, where the $H$-dependence of 
$\rho_{ab}$ becomes irreversible. 
Figure 3 shows the low-field MR term measured for $H$$\perp$$I$ 
(for clarity, the background MR, $\gamma H^2$, determined at high fields, 
is subtracted: 
$\Delta\rho_{ab}/\rho_{ab}$ = $(\Delta\rho_{ab}/\rho_{ab})^* + \gamma H^2)$. 
Initially the irreversibility appears as a small hysteresis on the MR curve; 
however, upon cooling to 10 K it becomes much more pronounced 
(the MR peaks are shifted from $H$=0 and strongly suppressed). 
We note that the first field sweep, which starts at 
$\Delta\rho_{ab}/\rho_{ab}$=0, differs significantly from the subsequent ones. 
The salient point here is that the resistivity does not return to 
its initial value after removing the magnetic field; 
hence, the system acquires a memory. 
In other words, the application of the magnetic field at low $T$ induces a 
persistent resistivity anisotropy in the CuO$_2$ planes.

\begin{figure}[bt]
\vspace{-30mm}
\includegraphics[scale=0.45]{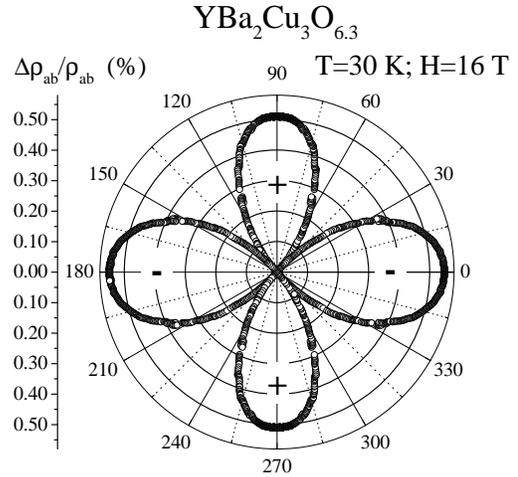}
\vspace{-40mm}
\caption{Dependence of the in-plane MR on the angle between $H$ and $I$ 
($H$$\parallel$$ab$; $H$=16 T). 
The sign of the MR is indicated.}  
\label{fig2} 
\end{figure}

The picture would be incomplete without the data on the $c$-axis 
transport (across the CuO$_2$ planes). 
It was shown that in antiferromagnetic YBCO below $T_N$, 
the suppression of spin fluctuations by the magnetic field 
results in a large positive out-of-plane MR \cite{MR}. 
Figure 4 shows an intriguing MR behavior produced by a 
superposition of the negative low-field MR feature on top of the 
large positive $\gamma H^2$ background in a sample with $T_N \ge 300$ K. 

\begin{figure}[bt]
\vspace{-0mm}
\includegraphics[scale=0.42]{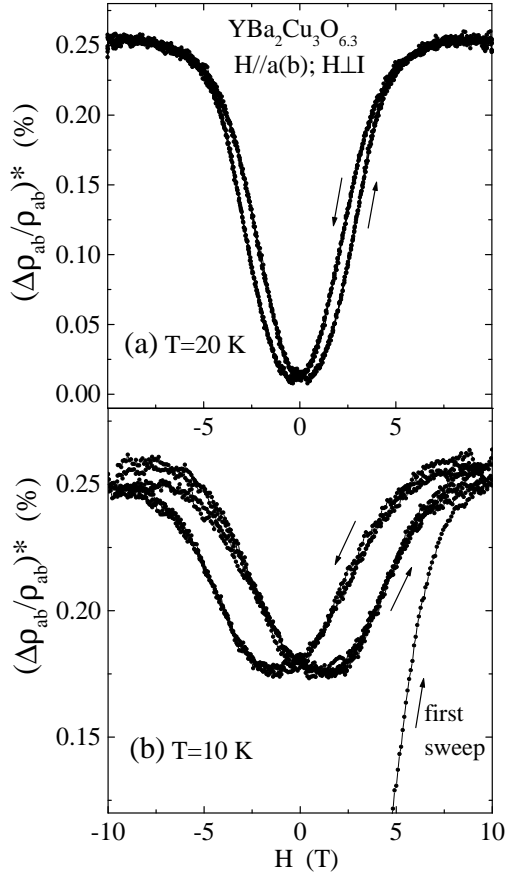}
\vspace{-15mm}
\caption{The low-field MR component. 
Each curve contains data of 4 field sweeps performed at a rate of 1 T/min. }  
\label{fig3} 
\end{figure}

\section{DISCUSSIONS}

It is very difficult to understand the MR anomalies presented here, 
especially the {\it memory effect}, without considering an 
inhomogeneous state or a superstructure in the CuO$_2$ planes 
instead of a uniform AF state. 
The picture of charged ``stripes" in the CuO$_2$ planes 
allows one to account for all the observed MR peculiarities, 
by assuming that the magnetic field gives rise to a directional ordering 
of the stripes \cite{stripe}. 
Actually, the aligning of stripes with confined carriers moving along 
would change the current paths and introduce the in-plane anisotropy. 
The rotation of stripes by the magnetic field gives a reasonable explanation 
for the in-plane MR with the $d$-wave-shaped angular dependence. 
Within this picture, the threshold field of several Tesla is presumably 
coming from the establishment of the directional order of the stripes. 

\begin{figure}[bt]
\vspace{-10mm}
\includegraphics[scale=0.4]{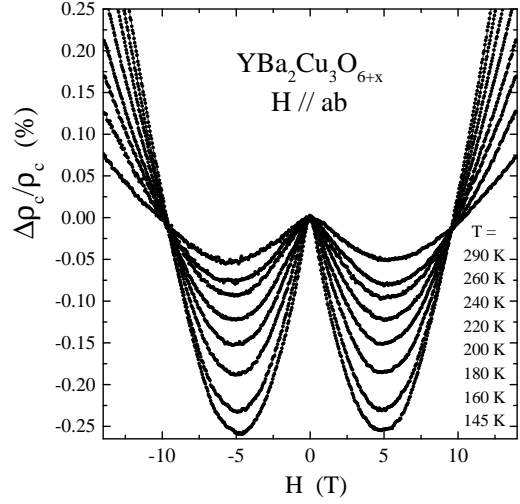}
\vspace{-48mm}
\caption{Transverse out-of-plane MR of YBCO in the 
AF composition, with $x \approx$ 0.28.}  
\label{fig4} 
\end{figure}

As the temperature is lowered, it is expected that the stripe dynamics 
slows down and the magnetic domain structure in the CuO$_2$ planes is frozen, 
forming a cluster spin glass. 
The spin-glass transition temperature has been reported to be about 20-25 K 
for the AF compositions \cite{Niedermayer}, 
which is in good agreement with the temperature 
where the hysteretic MR behavior is found. 

Though an explanation of the out-of-plane MR feature is not so straightforward, 
one can imagine that by adjusting the direction of the stripes 
in neighboring CuO$_2$ planes, the magnetic field increases the 
overlapping both between the stripes in the real space and between their 
quasi-1D carriers in the $k$-space, thereby enhancing the probability 
of the electron hopping. 

\begin{figure}[bt]
\vspace{-15mm}
\includegraphics[scale=0.43]{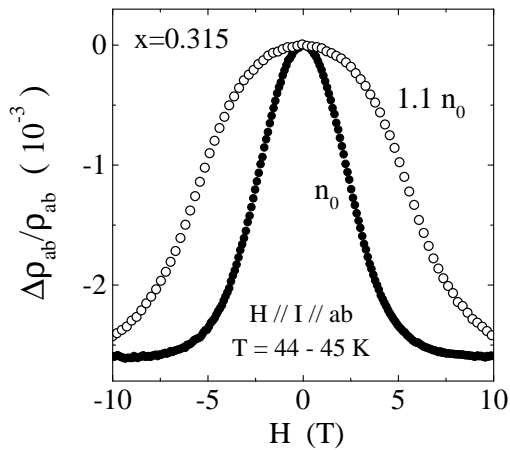}
\vspace{-58mm}
\caption{Longitudinal in-plane MR for two different carrier densities
measured on the same sample ($x$=0.315) before and after oxygen ordering. 
Solid circles show the data for the initial carrier density ($n_0$), and
the open circles show the data for $\sim$10\% larger carrier density.}  
\label{fig5} 
\end{figure}

One may wonder how the MR anomaly evolves when the carrier density
is increased, bringing the system to more metallic region.
Figure 5 shows the comparison of the MR data for two different 
carrier densities; the data were taken on the same sample, where the
$\sim$10\% increase in the carrier density was achieved by
keeping the sample at room temperature, which causes the oxygen
reordering.
It is clear that the threshold field for the stripe ordering 
increases with increasing carrier density.
The data in Fig. 5 suggest that the threshold field 
becomes inaccessibly high when the carrier density is increased to 
the superconducting region ($x>0.4$).
This is probably the reason why the MR anomaly reported here 
has never been observed in superconducting samples.

\section{CONCLUSION}

A variety of unusual MR features found in heavily underdoped YBCO have provided 
new information on the conducting charged stripes in the CuO$_2$ planes. 
The MR behavior implies that the stripes couple to the external magnetic field 
and undergo topological ordering at fields of the order of a few T,
although the actual mechanism that couples the stripes to the magnetic
field is not clear yet. 
Upon cooling the sample below $\sim$20 K, the dynamics of stripes slows down and 
the directional order of the stripes becomes persistent, 
giving rise to a ``memory effect" in the resistivity. 
These findings show that the magnetic field can be used as a tool 
to manipulate the striped structure and open a possibility to clarify 
the electron dynamics within the stripes.

\end{document}